\documentclass[sigconf]{acmart}

\def\BibTeX{{\rm B\kern-.05em{\sc i\kern-.025em b}\kern-.08emT\kern-.1667em\lower.7ex\hbox{E}\kern-.125emX}}
    
\copyrightyear{2019} 
\acmYear{2019} 
\setcopyright{acmlicensed}
\acmConference[IVA '19]{ACM International Conference on Intelligent Virtual Agents}{July 2--5, 2019}{PARIS, France}
\acmBooktitle{ACM International Conference on Intelligent Virtual Agents (IVA '19), July 2--5, 2019, PARIS, France}
\acmPrice{15.00}
\acmDOI{10.1145/3308532.3329473}
\acmISBN{978-1-4503-6672-4/19/07}
\begin{document}

\settopmatter{printacmref=true}
\fancyhead{}

%
\title{An End-to-End Conversational Style Matching Agent}

%
\author{Rens Hoegen}
\email{rhoegen@ict.usc.edu}
\affiliation{%
  \institution{University of Southern California}
  \city{Los Angeles}
  \state{California}
}
\author{Deepali Aneja}
\email{deepalia@cs.washington.edu}
\affiliation{%
  \institution{University of Washington}
  \city{Seattle}
  \state{Washington}
}
\author{Daniel McDuff}
\email{damcduff@microsoft.com}
\affiliation{%
  \institution{Microsoft Research}
  \city{Redmond}
  \state{Washington}
}
\author{Mary Czerwinski}
\email{marycz@microsoft.com}
\affiliation{%
  \institution{Microsoft Research}
  \city{Redmond}
  \state{Washington}
}

%
\renewcommand{\shortauthors}{Hoegen, et al.}

%
\begin{abstract}
We present an end-to-end voice-based conversational agent that is able to engage in naturalistic multi-turn dialogue and align with the interlocutor's conversational style. The system uses a series of deep neural network components for speech recognition, dialogue generation, prosodic analysis and speech synthesis to generate language and prosodic expression with qualities that match those of the user. We conducted a user study (N=30) in which participants talked with the agent for 15 to 20 minutes, resulting in over 8 hours of natural interaction data. Users with high consideration conversational styles reported the agent to be more trustworthy when it matched their conversational style. Whereas, users with high involvement conversational styles were indifferent. Finally, we provide design guidelines for multi-turn dialogue interactions using conversational style adaptation. 
\end{abstract}

%
%


%

%

%
\maketitle

\section{Introduction}
Personal assistants (e.g., Alexa, Siri, Cortana, and Google Now) and ``bots'' interacting in natural language (e.g., Messenger, Skype, and Sina Weibo), have created new platforms for human-computer interaction. In the U.S. nearly 50 million (or 1 in 5) adults are estimated to have access to a smart speaker (e.g., Amazon Echo) for which voice is the primary interface. Many more have access to an assistant on their smartphone or smartwatch.

Despite the prevalence of these devices and the considerable investment in research and development devoted to them, it is still not very natural to engage with AI agents in extended interactions and open-ended conversations. In part, this is because the agents are unable to have multi-turn conversations with the users and fail to adapt to the social behaviors of humans, although some work on improving interactions has been done with embodied agents~\cite{el2016towards,gratch2007creating}. This large gulf in expectations is perhaps part of the reason why conversational agents are only used for very simple tasks and often disappoint users~\cite{luger2016like}. 

It has been shown that rapport and information disclosure are higher with assistants that can engage in social dialogue and respond appropriately~\cite{bickmore2000weather,bickmore2001relational}. REA (an embodied real-estate agent)~\cite{bickmore2000weather}, SARA (socially-aware robot assistant)~\cite{matsuyama2016socially} and SimSensei (an embodied conversational agent used for healthcare interviews)~\cite{devault2014simsensei}, are some examples of conversational agent systems that attempt to leverage social dialogue to establish rapport with their users.

While agents' ability to engage in dialogue has been studied quite extensively~\cite{bickmore2000weather,traum2002embodied}, the conversational style of these agents has received less attention. However, it has been shown that people's perceptions of conversational agents are influenced by the interaction style of the agent~\cite{bickmore2005social,matsuyama2016socially,pecune2018field}. 

In terms of understanding human-to-human conversation, Deborah Tannen's theory on conversational style is the most widely used~\cite{tannen87style,tannen05style}. Tannen defines style as ``...the use of specific linguistic devices, chosen by references to broad operating principles or conversational strategies.'' Tannen's conversational style framework categorizes styles on an axis ranging from High Consideration (HC) to High Involvement (HI). An HC interaction style emphasizes consideration and independence.  It is characterized by longer pauses and hesitations and the use of moderate paralinguistic characteristics. An HI interaction style, on the other hand, emphasizes interpersonal involvement and interest. It is characterized by speaking faster and more loudly, overlapping with the other speaker, and with fewer pauses between conversational turns. 

Niederhoffer and Pennebaker have done studies that show that people align their linguistic style in social interactions~\cite{niederhoffer2002linguistic}. Linguistic mimicry, the adoption of another's language style, is a behavior that has social benefits~\cite{otterbacher2017show} where greater empathy is associated with mimicry in conversations. Therefore, designers of an artificially empathetic conversational agent explicitly identified conversational style matching as a desirable feature; however, it was not included in their published design~\cite{morris2018towards}. In other words, style matching in conversational agent design warrants further exploration.

Conversational style includes prosody, word choice, and timing, for example. We distinguish style (the ``how''), from any topical information transferred (the ``what''); we can provide the same information with different styles. Prior work defined conversational style parameters amenable to computation and that were able to resemble style matching mechanisms between humans~\cite{thomas2018style}, and we were influenced by this work. The purpose of the current study was to understand if an intelligent, conversational agent could perform conversational style matching ``on-the-fly'' and, if so, how this affects the perception of such an agent. 

Designing an end-to-end conversational agent is complex. It requires speech recognition, dialogue generation, and speech synthesis. Furthermore, all these need to operate in real-time, without any unnatural delay in the timing of responses between the agent and the human. We designed a conversational agent that is capable of multi-turn dialogue without human intervention. Via a generative neural language model, the agent is capable of relatively open-ended discourse and can respond to utterances from the interlocutor in an understandable way. By combining this with automatic speech and paralinguistic sensing capabilities, and prosodic control of the agent's synthesized speech, we created a novel conversational agent system that is capable of real-time style matching. 

The contributions of the paper are as follows: 1) we present an end-to-end automated conversational agent capable of conversational style matching, 2) we conducted a user study to examine the effect of conversational style matching on perception where we showed that conversational style increases a user's trust in an agent, compared to an agent without conversational style matching, 3) we provide design guidelines for conversational systems with conversational style matching.

\section{Related Work}
\subsection{Conversational style}
Most of the related research on conversational style matching is based upon the work of Tannen, who described several markers that define conversational style~\cite{tannen05style}. These markers include: Topic (e.g., personal pronoun usage, persistence), Pace (e.g., speech rate, pauses), Expressive Paralinguistics (e.g., pitch and loudness shifts) and Genre (e.g., story framing). Based on the usage of these stylistic markers, people can be placed on an axis ranging from HC to HI.

In more recent work, Shamekhi et al. looked at conversational style in human-agent interactions~\cite{shamekhi2016exploratory}. They examined whether there was a particular conversational style (HI or HC) that was preferred for a conversational agent in a structured conversation. However, rather than finding one specific style that worked best, they found that participants liked whichever agent matched their own conversational style.

Thomas et al. looked at the conversational style of participants while performing a simple information seeking task~\cite{thomas2017misc,thomas2018style}. This task was similarly structured to how a person might use an intelligent agent such as Cortana or Siri. They found that even in these tasks, participants aligned their conversational styles over time. It took less effort to complete tasks for participants who aligned their style compared to those who did not. 

Conversational style matching can be seen as a type of entrainment, as it relates to people synchronizing their behavior within an interaction~\cite{hirschberg2011speaking,bevnuvs2014conversational}. Entrainment has been used in related work to generate more accurate and realistic reactions~\cite{jauk2011dynamic} or backchannels~\cite{inden2013timing}.
Levitan et al. implemented acoustic-prosodic entrainment in a conversational avatar and observed an increase in perceived reliability and likability of the system~\cite{levitan2016implementing}. In a text-chat interface, Scissors et al.~\cite{scissors2008linguistic,scissors2009cmc} found that lexical entrainment was associated with trust between partners. Specifically, pairs with high trust exhibited greater repetition of shared terms than did pairs with lower trust~\cite{scissors2008linguistic}. Subsequent work found that this was driven primarily by certain types of terms (e.g., those with greater positive emotional valence), and that not all similarity increased trust. The similarity in negative terms was associated with decreased trust~\cite{scissors2009cmc}. We were inspired by this work and wanted to build upon it using an automated agent capable of voice-based interactions.


\subsection{Agents and Trust}
Conversational Agents have been studied extensively, with trustworthiness being an important aspect of evaluating and designing these systems~\cite{isbister2002design,ruttkay2006brows}. Trust has been mentioned as an important goal for a socially intelligent agent. For example, Elofson et al.~\cite{elofson2001developing} investigated the role intelligent agents can play in the development of trust within organizations. 
Cassell and Bickmore argue that the methods that humans use for establishing trust can also be used by intelligent agents~\cite{cassell2003negotiated}. As shown in their studies with the embodied agent REA, by incorporating small talk, REA was perceived as more trustworthy by extroverted participants than an agent that does not engage in this social behavior~\cite{bickmore2001relational}. 

In a field trial with the SARA agent, Pecune et al. found that the interactional features (e.g., frequency of interruptions and number of turns) played a role fostering both rapport and task performance. The agent's social awareness was especially important due to the short duration of the interactions, as there is likely a social politeness boundary that should not be broken in short interactions~\cite{pecune2018field}. Rapport is characterized by a close and harmonious relationship, and one would think that this quality would also strengthen trust between an agent and a human.

In a study, Gratch et al. found that participants disclosed more to SimSensei, a virtual interviewer for healthcare when presented as an artificial intelligence, rather than an avatar controlled by a human operator in a medical interview~\cite{gratch2014s}. These results suggest that an intelligent virtual agent might be used to obtain more truthful information within medical domains.


\begin{figure*}
\includegraphics[width=1\linewidth]{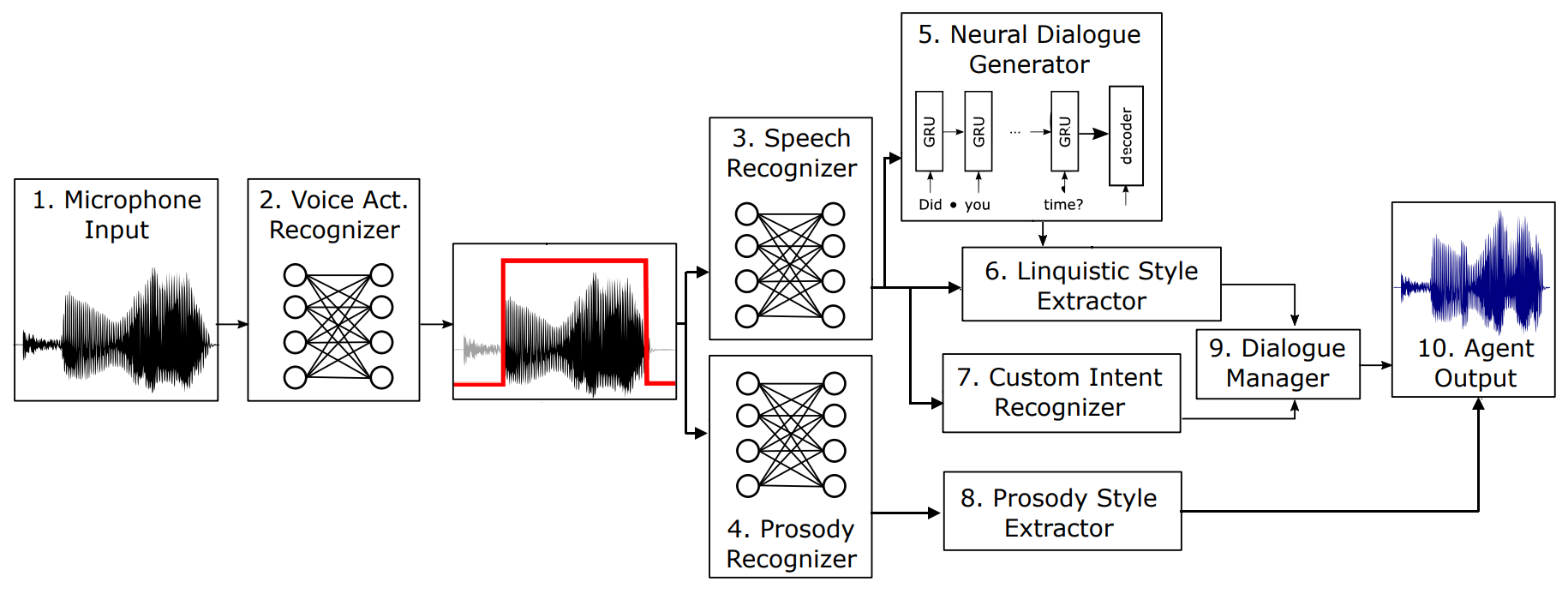}
\caption{Architecture of our dialogue agent. The microphone input is first processed to extract voiced segments.  The voiced segments are then passed to neural speech recognition and voice prosody recognition.  The text output from the speech recognition was input to a neural dialogue generation model and also to the LUIS intent recognition engine. The recognized speech and generated dialogue response were both sent to the conversational style extractor. Finally, a rule-based dialogue manager selected the utterance to synthesize.}
\label{fig:architechture}
\end{figure*}

\section{Hypotheses}
For this study, our goal was to build upon work on conversational style matching using a fully automated, voice-based interface. Based on the work of Thomas et al.~\cite{thomas2018style}, we know that people slowly align their conversational style while performing a task. Secondly, based on the work of Shamekhi et al.~\cite{shamekhi2016exploratory}, we have seen that people prefer a conversational agent that matches their own conversational style.

Bolstered by these studies, we desired to verify whether an agent that performs conversational style alignment is preferred to one that does not. We expected participants to prefer this style matching behavior. Additionally, there is some evidence that HC participants are more likely to adjust their range, pitch, and intonation with their partner. We, therefore, suspect that style matching of an agent might be more effective for participants with an HC style~\cite{shamekhi2016exploratory}. 

Additionally, Bickmore and Cassell have shown that an agent's perceived trustworthiness can be positively affected by having the agent perform the social dialogue in the form of small talk~\cite{bickmore2001relational}. We investigate whether conversational style matching further improves this effect. Therefore, our agent uses social dialogue in its interactions, and we measure the perceived trustworthiness of the agent. As such, our hypotheses for this study are as follows:
\\~\\
\textbf{H1a.} People prefer interacting with an agent that uses conversational style matching, as opposed to one that does not.
\\
\textbf{H1b.} People trust an agent using conversational style matching more than one that does not.
\\
\textbf{H2.} Preferences about an agent depend on their conversational style. In particular, people with an HC style prefer style matching.
\\~\\
In order to investigate these research questions, we constructed an intelligent agent that is capable of conducting open-ended conversations. This agent does not require any intervention of a human in order to converse with people, it responds to user input fully automatically. By using speech recognition and paralinguistic parameter recognition, we can apply conversational style matching in real time. We designed an experiment where participants spoke to the agent for about 15 minutes. Participants would either interact with an agent that applied conversational style matching or an agent that did not. Results from this study enabled us to formulate design guidelines for conversational agents, based on our findings.

\section{Intelligent Agent Design}
We built an intelligent conversational agent capable of conducting an open-ended conversation with a person. The agent was built using the open source Platform for Situated Interaction (PSI)~\cite{bohus2017rapid}. Figure~\ref{fig:architechture} shows the architecture of the agent and below we describe the constituent components.

\subsection{Sensing}
\emph{\textbf{Speech Recognition:}}
Voice activity was detected with the Windows system voice activity detector (Microsoft, Inc.). Voiced audio was then passed to the Bing speech recognition engine\footnote{https://azure.microsoft.com/en-us/services/cognitive-services/speech/} which returned the recognized textual string.

\emph{\textbf{Paralinguistic Parameter Recognition:}}
For the voiced audio segments, the paralinguistic speech features were also extracted. Specifically, we detected the fundamental frequency or pitch ($\textit{f}_0$), and the root mean squared (RMS) energy reflecting the loudness of the voice. These features were extracted using a digital signal processing approach similar to that described by Han et al.~\cite{han2014speech}. 

\subsection{Synthesis}
\emph{\textbf{Dialogue Generation:}}
We used a data-driven paradigm of conversation generation~\cite{sordoni2015} in which neural models are built, typically from large-scale unconstrained datasets. Our model was trained on social media data that were mined from the Twitter fire hose. The neural models can then return a ranked list of the most likely responses for any text-based input. Despite the diverse and relatively unconstrained dialogue examples, the model is able to learn to conduct simple conversations. However, while these unconstrained models are able to conduct general chit-chat effectively, they are not goal driven, nor do they have a deep knowledge-base about the world. Therefore, we added the capability to identify specific intents in text to our system by using the Language Understanding and Intent Service (LUIS)\footnote{https://www.luis.ai/home}. LUIS can recognize intents by supplying training examples of them, after which it will be able to detect variations of the intents. An example of an intent might be ``Movies'' and a training example might be ``Do you like movies?'' or ``What are your favorite films?". LUIS was trained to recognize 10 to 15 specific intents (165 intents total across all tasks). When an intent is recognized by the system, the system will respond with a predefined scripted response using it dialogue manager. These responses are less open-ended and will often pose a specific question to the user, allowing us to extend the interaction time participants had with the agent.

\begin{figure}
\newcommand{\UU}{\textbf{User:}}
\newcommand{\II}{\textbf{Agent:}}
\centering


\sffamily\small
	\begin{flushleft}\UU\ Do you know if there are any museums around Buckingham Palace in London? \\
	\II\ I am sure there will be some in the area.\\
	\UU\ What about parks?\\
	\II\ \color{blue}I know Buckingham Palace is surrounded by some beautiful parks and you can checkout the palace. Or you could go to Hyde Park where you can row a boat or go to the Rose Gardens. What do you think?\\
    \color{black}
	\UU\ I think that sounds nice... I think I'd also like to go for afternoon cream tea... \\
    \hspace{1cm} \textit{Agent interrupts} \\
	\II\ Yes, it does. \\
    \UU\ ...Do you know anywhere that does that? \\
	\II\ I believe I don't know of any. \\
    \UU\ You don't know of any? \\
	\II\ No, I don't sorry. \end{flushleft}


\caption{Transcript of an interaction where the user plans a trip to London.}
\label{fig:conversation_examples}
\end{figure}

Figure~\ref{fig:conversation_examples} shows an excerpt from a conversation with the agent collected during our user study.  The dialogue generated by the neural model is shown in black, and the scripted responses to detected intents are highlighted in blue. A majority of the utterances in each conversation were generated by the neural dialogue model.

\emph{\textbf{Speech Synthesis:}}
We use the speech synthesis markup language (SSML)~\cite{burnett2004speech} to vary the prosodic qualities of the generated speech (pitch, loudness and speech rate). Within our system, SSML allows each of these properties to be specified at the utterance level. For both pitch and loudness, the system allows variation between five different levels (ranging from very low pitch/loudness to very high pitch/loudness). Speech rate is changed with a floating-point number, where 1 represents standard speed, 2 is double speed, 0.5 is half speed and so forth.

\section{Conversational Style Matching}
In order to match the participant's conversational style, the agent leverages its sensing components (speech recognition and the paralinguistic parameter recognition). Then, based on the data collected through these components, the agent calculates several variables that define the user's conversational style. These variables are based on the work of Thomas et al. and have been shown to capture the consideration-involvement dimension of conversational style~\cite{thomas2018style}. The specific features used in this study were selected to be computable at scale in real time, without intervention. By manipulating the variables in its own responses, the agent performs conversational style matching. The agent matches both the content of its utterances, as well as the prosodic qualities of the spoken response.

\subsection{Content Variables}
We used four different variables to define a participant's word usage: one variable tracks pronoun usage, two variables measure repetition and one measures utterance length.

\emph{\textbf{Pronoun use:}}
The first variable is personal pronoun usage, which measures the rate participants use personal pronouns in their speech (e.g. you, he, she, etc.). This measure is calculated by calculating the usage ratio of these words compared to other words occurring in each utterance.

\emph{\textbf{Repetition:}}
In order to measure repetition, we used two variables that both relate to the repetition of terms (A term in this context is a word that is not considered a stop word). Repetition can be seen as a measure of persistence in introducing a specific topic. The first of the variables measures the rate of repeated terms per individual utterance. The second is a ratio of sentences that contained any repeated terms.

\emph{\textbf{Utterance length:}}
The last content variable is a simple measure of the average number of words per utterance and represents the participant's average utterance length.

\subsection{Acoustic Variables}
We used three different variables that describe the participant's paralinguistic features in their speech; speech rate, pitch variance, and loudness variance.

\emph{\textbf{Speech rate:}}
The speech rate variable measures how quickly participants speak. As such, it is somewhat related to the utterance length to define verbosity. Speech rate measures how many words the participant says per second. This is calculated by using the Windows system voice activity detector to determine whether the participant is speaking and then comparing the recognized text with the duration of the voice activity, giving us the average number of words per second for each utterance. 

\emph{\textbf{Pitch:}}
The pitch feature was measured by determining the participant's $\textit{f}_0$ on a per utterance basis and was stored as such.

\emph{\textbf{Loudness:}}
The final feature, loudness, was measured in a similar way as to how Pitch was measure--by determining the detected RMS energy of each utterance.

The above variables were all calculated in real time while the participant was interacting with the agent.

\subsection{Conversational Style Matching Synthesis}
In order to perform conversational style matching, the agent matches the participant on the previously defined conversational style variables. There are two ways the agent does this. For the word choice and utterance length, the agent adjusts the content of the utterance in order to match the participant more closely. For the prosody variables and speech rate, the agent changes the prosody and pace of its own responses.  In all cases, the conversational style variables were aggregated over the last five utterances to reduce the likelihood of dramatic changes in style from one utterance to the next.

\emph{\textbf{Utterance content matching:}}
In order to match the conversational style to the utterance content, the agent adjusts the ranking order of the responses generated by our dialogue model. This is done by calculating the conversational style variables (word choice and utterance length) of the top 10 responses returned by the dialogue model's beam search algorithm. The answers are then re-ranked based on how closely they match the content variables of the interlocutor. Since the top 10 responses of the dialogue model are generally very close in meaning to each other, we can adapt the conversational style of the response to that of the user, without changing the meaning of the response.  

\emph{\textbf{Acoustic matching:}}
The agent matches the participant's prosody and pace using SSML to adjust the pitch, loudness and speech rate of its utterances. The matching of prosody and pace is done on a local level, meaning the agent matches the participant's changes, rather than match their overall prosody and pace. In order to do this, the agent tracks the average pitch and loudness of the participant overall and uses this as a baseline of the participant's speech. It then determines the current style by comparing the values of the last five utterances to this baseline. After establishing a baseline, the agent matches any detected increase or decrease in these values. Because of this approach, the agent requires a stable baseline of several utterances before it can perform matching. As such, it does not perform prosody matching on the first four utterances.

\section{User Study}
Our user study protocol was approved by the Institutional Review Board (IRB) of Microsoft. The study used a between subjects' design. Participants either interacted with an agent using conversational style matching, or one that did not. In order to reach 15 minutes of interaction time with the agent, we defined a set of tasks (scenarios) that the participant would discuss with the agent (we define the specific tasks further in the Tasks subsection). After the interaction with the agent, participants filled out several surveys used to investigate our hypotheses (the specific surveys are defined in the Materials subsection). Participants would take approximately 15 minutes to fill out the surveys, resulting in a total time of 30 minutes to complete the study.

\begin{table}
  \caption{Summary of Participation}
  \label{tab:participants}
  \begin{tabular}{rcc}
    \toprule
     & Control & Matching \\
    \midrule
    Gender & M = 8, F = 7 & M = 8, F = 7\\
    Conv. Style & HC = 5, HI = 10 & HC = 9, HI = 6 \\
    Age & $\mu$=33 ($\sigma$=10.9) & $\mu$=31 ($\sigma$=6.1) \\
  \bottomrule
\end{tabular}
\end{table}

\subsection{Participants}
We recruited 30 participants (14 females) total. The average age was 32 years (SD = 8.77). Although 16 participants were identified as HI speakers and 14 were HC, we were unfortunately not able to balance the study for conversational style as this was a self-reported measure collected during the study. The participants reported varying levels of familiarity with speech-based interfaces. Twenty-two participants owned a smart speaker. 
Table~\ref{tab:participants} contains a summary of the participants.

\subsection{Apparatus}
\begin{figure}
\includegraphics[width=\linewidth]{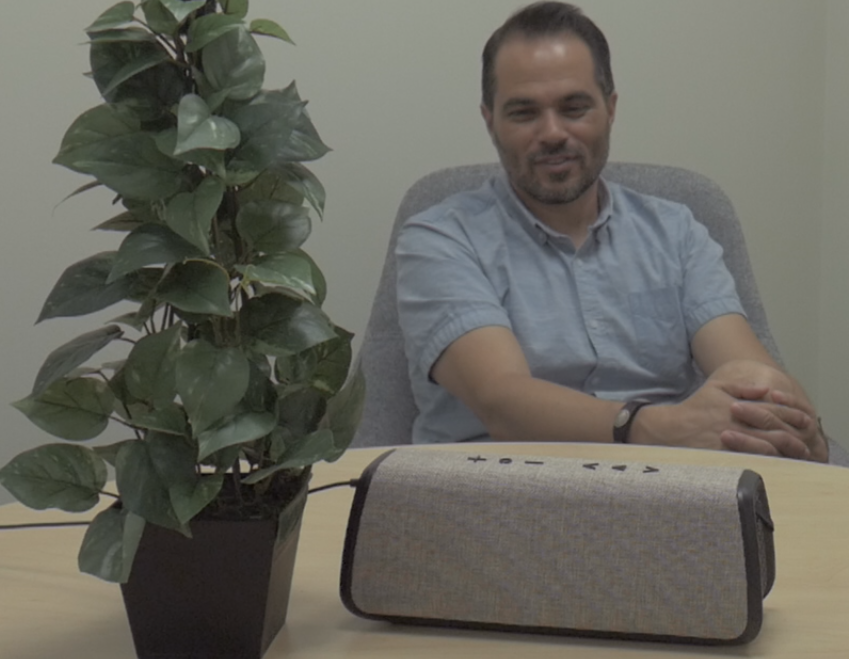}
\caption{Setup for the user study}
\label{fig:apparatus}
\end{figure}

During the user study, the participants were seated singly in a room.  A speaker (FUGOO Style-S Portable Bluetooth Speaker) was placed on a table opposite them.  A microphone was placed on a table next to the participant. The configuration of the apparatus can be seen in Figure~\ref{fig:apparatus}.

\subsection{Materials}
We used several self-report surveys instruments to capture the experience of the users. 
We first captured several participant traits:

We recorded the participants' self-reported conversational style using a survey created by Shamekhi et al.~\cite{shamekhi2016exploratory}, based on Tannen's~\cite{tannen05style} style theory.
We captured the personality of participants using the Big Five Personality Inventory survey by John et al.~\cite{john1999big}.

The second set of surveys were used to capture the participants' attitude towards the agent and ratings of their interaction with it.

In order to measure how participants experienced the interaction with the conversational agent, we used the same questionnaire as Shamekhi et al.~\cite{shamekhi2016exploratory}. The interaction questionnaire contains several questions relating to the participant's overall impression of the interaction, as well as some questions on the perceived trustworthiness of the agent. We, therefore, created two composite variables, one relating to interaction and one on the trust rating of the agent.

In order to capture the participants' impressions of the agent, we used the Godspeed questionnaire by Bartneck et al. ~\cite{bartneck2009measurement}. This survey was originally used in robotics, therefore some modifications were made (e.g., we excluded questions on physical movement).

\begin{figure*}
\includegraphics[width=\linewidth]{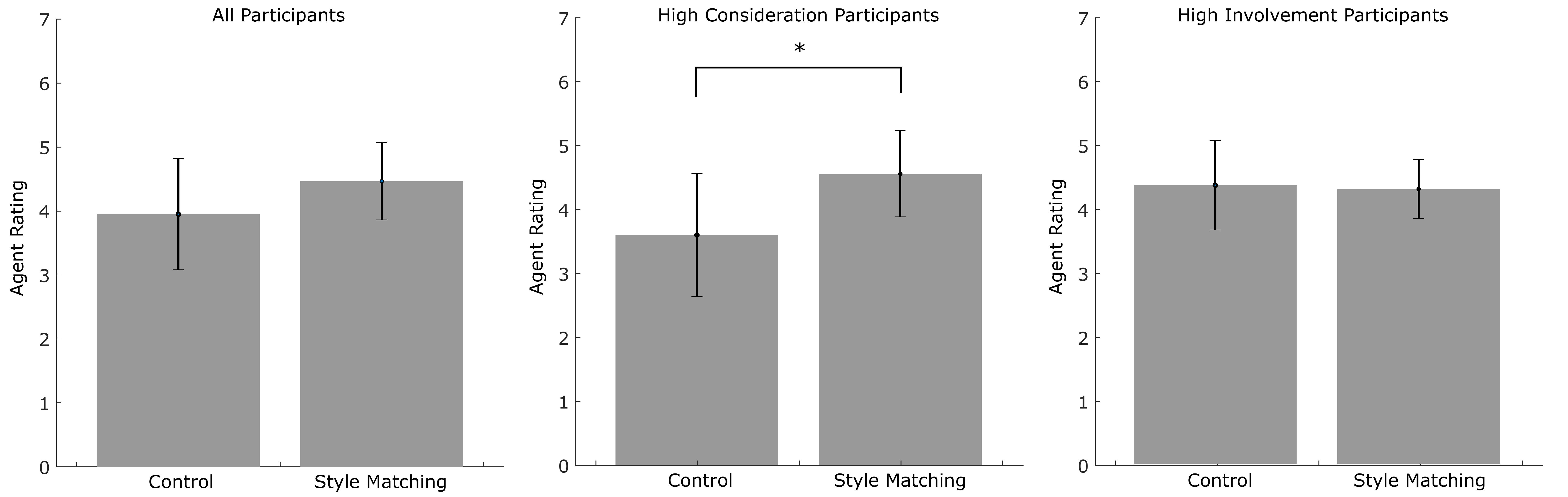}
\caption{Left: Rating of trust in the agent for all subjects (N$_{control}$ = 15, N$_{matching}$ = 15). Middle: Rating of trust in agent for high consideration subjects (N$_{c}$ = 5, N$_{m}$ = 9). Right: Rating of trust in agent for HC subjects (N$_{c}$ = 9, N$_{m}$ = 6).}
\label{fig:Results}
\end{figure*}

\subsection{Tasks}
We defined several tasks for participants to perform during the study. Each task described a scenario that the participant discussed with the agent. For each task, 10 to 15 scripted responses were created, and these responses were triggered when the LUIS intent recognition engine detected a specific intent relating to one of the scripted responses. (see Figure ~\ref{fig:conversation_examples} for an example of a transcript).

The participants performed the following tasks: In a simple introduction task, participants scheduled a lunch meeting with the agent. In the second task, participants were invited to talk about their personal life (e.g., hobbies, work). In the third task, participants planned a trip to London with the agent. The fourth task involved organizing a party, and the fifth and final task had participants talk about movies with the agent.






\section{Results}
Overall, the participants spend at least three minutes in each of the five tasks and spoke with the agent for an average of 18 ($\sigma$ = 1.94) minutes.  This resulted in close to nine hours of interaction with the agent across the 30 participants.

We first analyzed the results of the participants' ratings of the overall interaction (H1a) and the perceived trustworthiness (H1b) of the agent, both composite measures obtained through the interaction survey. We did not find a significant difference between the style matching agent and the control condition for either overall interaction (t(28) = .21, p = .837), nor for the trustworthiness rating (t(28) = 1.26, p = .22).

Following this, we performed a set of two-way ANOVAs on the effects of the experimental condition (conversational style matching or control) and the participants' conversational style (HC or HI), for both the interaction and the trustworthiness rating of the agent. For the first ANOVA on the interaction rating, there was no significant interaction between condition and the participants' conversational style (F(1, 26) = .001, p = .98). So we cannot confirm hypothesis 1a.

The second two-way ANOVA on the agent's trustworthiness ratings revealed that there was a trend of an interaction between the condition and the participants' conversational style (F(1, 26) = 3.347, p = .079). 
We investigated this further with a simple effects analysis, which showed that this interaction trend appears to be driven by participants with an HC conversation style (p = .027). HC participants rated the conversational style matching agent more trustworthy ($\mu$ = 4.56, $\sigma$ = .24) than the control condition ($\mu$ = 3.6, $\sigma$ = .33). However, for HI participants there was no difference (p = .875), as they rated the style matching agent ($\mu$ = 4.3, $\sigma$ = .3) similar to the control condition ($\mu$ = 4.36, $\sigma$ = .23). Thus, showing that the agent's conversational style matching appears to be more important for HC participants. This result provides some support for hypotheses 1b and 2. Figure~\ref{fig:Results} shows the mean agent ratings for the trustworthiness scores for the two experimental conditions. On the left, we show the results across all participants, in the middle the results for the participants with high consideration conversational styles and on the right the results for the participants with high involvement conversational style. 

Lastly, since we also measured personality with the big-5 questionnaire, we investigated whether there is a correlation between personality and trust. However, we found no significant correlations, with the highest correlation for the personality types being between extraversion and trust (r=.27, p=.14). 



\section{Discussion}
Using our conversational agent and dialogue model, the participants were able to sustain conversations for the expected 15 minutes. Scripting a natural dialogue of this length was novel and challenging. However, the neural language model was able to successfully generate dialogue turns that were appropriate for general chit-chat, while the scripted responses were successful in responding to on-topic utterances and drive the conversation.

We found that participants with an HC conversational style rated the agent as more trustworthy when it matched their conversational style, compared to when it did not. Whereas, participants with an HI conversational style rated both agents similarly. We did not find any differences in the rating of interaction quality. However, the small number of participants in our study could possibly limit the significance of our results. We ran 30 participants total spread across 2 conditions. As a comparison, Niederhoffer and Pennebaker ran a study on linguistic style matching between dyads with a total of 62 dyads~\cite{niederhoffer2002linguistic}. It is expected that the effect sizes of subtle adaptations in conversational style might be small after a 15-minute interaction. Nevertheless, it could remain a very important design consideration when a system is deployed longitudinally ``in-the-wild''.

One possible explanation for the result of HC users rating a conversational style matching agent as more trustworthy might lie in how these participants are more considerate of their partner while having a conversation. As such, the agent adapting to the participant's conversational style might be appreciated more by an HC speaker than by an HI speaker, thus resulting in a higher trustworthiness rating. To put it simply, HI speakers might be less troubled by an interlocutor that does not match their style. 

We observed some differences in how participants interacted with the agent. Some of these differences might have affected how participants experienced the agent (regardless of condition). For example, for 11 out of the 30 participants some speech overlap occurred with the agent (see figure ~\ref{fig:conversation_examples} for example). Overlap can be seen as an aspect of an HI conversational style, but in our study, this was not part of the design and occasionally occurred in both conditions. There are several reasons why speech overlap occurred. For example, it might be because of a participant's conversational style (i.e., HI participants are more likely to overlap in speech). However, it also happened because of limitations in the design of the system, as the agent responded to every single utterance it recorded. At times, the agent made segmentation errors by splitting up a single utterance into multiple, and as such would respond to several utterances in a row. If a participant was not aware of this, they might consider that the agent was finished speaking after the first utterance, and then overlap with the agent's response to the second part of the utterance. Secondly, if these participants did not finish their sentence, the agent would still respond to what it heard. This behavior was not expected by some participants and could also occasionally lead to nonsensical responses of the agent. In general, participants that took longer pauses while interacting with the agent did not experience overlap as often, whereas those that spoke faster would often overlap with the agent (possibly thinking that the agent was finished speaking when it had not). Some of these types of overlap observed could, unfortunately, go on for several rounds of the conversation. 

\section{Design Guidelines}
We define some design recommendations to keep in mind when developing a conversational style matching agent using unconstrained dialogue models. The guidelines are based on our observations of participants interacting with the conversational agent.

In order for conversational style matching to be most effective, some care needs to be taken. One major pitfall is that the positive effect of the matching condition could easily be overshadowed and undone by the negative effects of other limitations of the system. In some cases, the matching could even exacerbate these negative effects. For example, if an agent gives answers that frustrate its user, this could influence the user's conversational style. This frustration is then matched by the agent, which will most likely affect the experience of the participant. This could create a negative spiral of negative style matching. As such, it is important to consider this while designing a system that uses style matching.

Keeping the above in mind, if the system does have some significant flaws, it would be wise to notify the participant of this and not oversell the system's capabilities. For example, if it takes a while for a system to process and answer specific questions and style match, the user needs to be made aware of this. If the participant is informed of this before the interaction, it could create some understanding and perhaps reduce frustration from having to wait for an answer. We have observed this in subsequent studies which we will report on in the future.

As previously noted, there can be a lot of overlap between the temporal aspects of the speech of the agent and the participant. Although overlap in speech can be common in human-human dialogue, this is usually an aspect of an HI conversational style. Therefore, in order to build an HC-appropriate conversational agent, this needs to be considered. There are several, fairly simple changes that could deal with these types of issues. One solution is allowing the agent to be interrupted (i.e., the agent stops talking when overlapping speech is detected), thus giving the floor to the participant to continue speaking. Another way of avoiding overlap is to filter out stop words and interjections from the participant. Currently, the system does not do this and therefore responds to any utterance of the participant, which can potentially lead to some confusion on the participant's side, since people often do not even realize they are using stop words or are interjecting, as we observed.

One other useful addition could be the incorporation of more frequent usage of pauses by the agent. This was not implemented in the current system, but there is a great difference in usage of pauses between HI and HC speakers and this might have a big impact on the participant's perception of the system.

\section{Conclusions}
We designed an end-to-end voice-based conversational agent capable of multi-turn dialogue to study the effect of conversational style matching on participants' perception of the agent. Thirty participants interacted with the agent for approximately 15 minutes. We found that individuals with High Consideration conversational styles were more likely to trust an agent that matched their conversational style, whereas those with the High Involvement style were indifferent. Despite the subtle nature of conversation style matching, we were able to observe an effect on participants' trust ratings after only 15 minutes of agent interaction. This is very encouraging for future agent designs that aim to leverage subtle conversational matching (or entrainment).

Future work will consider the use of an end-to-end conversational agent that combines visual information with textual inputs to generate image-grounded dialogue~\cite{mostafazadeh2017image,huber2018emotional,huber2018facial}. Previous work has shown that image-grounded models allow for more appropriate dialogue generation. We also aim to deploy an agent longitudinally to measure the effects of style matching over longer interactions and in more natural contexts.

%
\bibliographystyle{ACM-Reference-Format}
\balance
\bibliography{bibliography}

\end{document}